\documentclass[USenglish,autoref,pdfa,a4paper,authorcolumns]{lipics-v2021}
\usepackage{amsbsy}
\usepackage{amstext}
\usepackage{stmaryrd}
\ifx\hypersetup\undefined
  \AtBeginDocument{%
    \hypersetup{unicode=true,pdfusetitle,
 bookmarks=true,bookmarksnumbered=false,bookmarksopen=false,
 breaklinks=false,pdfborder={0 0 0},pdfborderstyle={},backref=false,colorlinks=false}
  }
\else
  \hypersetup{unicode=true,pdfusetitle,
 bookmarks=true,bookmarksnumbered=false,bookmarksopen=false,
 breaklinks=false,pdfborder={0 0 0},pdfborderstyle={},backref=false,colorlinks=false}
\fi

\makeatletter

\theoremstyle{plain}
\newtheorem{thm}{\protect\theoremname}
\theoremstyle{definition}
\newtheorem{example}[thm]{\protect\examplename}
\theoremstyle{definition}
\newtheorem{defn}[thm]{\protect\definitionname}
\theoremstyle{plain}
\newtheorem{lem}[thm]{\protect\lemmaname}
\theoremstyle{plain}
\newtheorem{cor}[thm]{\protect\corollaryname}

\nolinenumbers
\author{Xiang Long}{Cornell University, Ithaca, New York, USA}{xiang@cs.cornell.edu}{}{}
\authorrunning{X. Long}
\Copyright{Xiang Long}

\usepackage[misc]{ifsym}
\usepackage{tikz-cd}
\newcommand{\llb}{\llbracket}
\newcommand{\rrb}{\rrbracket}
\newcommand{\llp}{\llparenthesis}
\newcommand{\rrp}{\rrparenthesis}
\newcommand\barnone{\bar{\rule{0mm}{1ex}\hspace{1ex}}}

\makeatother

\providecommand{\corollaryname}{Corollary}
\providecommand{\definitionname}{Definition}
\providecommand{\examplename}{Example}
\providecommand{\lemmaname}{Lemma}
\providecommand{\theoremname}{Theorem}

\begin{document}
\title{MatchKAT: An Algebraic Foundation For Match-Action}

\maketitle
\begin{abstract}We present MatchKAT, an algebraic language for modeling
match-action packet processing in network switches. Although the match-action
paradigm has remained a popular low-level programming model for specifying
packet forwarding behavior, little has been done towards giving it
formal semantics. With MatchKAT, we hope to embark on the first steps
in exploring how network programs compiled to match-action rules can
be reasoned about formally in a reliable, algebraic way. In this paper,
we give details of MatchKAT and its metatheory, as well as a formal
treatment of match expressions on binary strings that form the basis
of ``match'' in match-action. Through a correspondence with NetKAT,
we show that MatchKAT's equational theory is sound and complete with
regards to a similar packet filtering semantics. We also demonstrate
the complexity of deciding equivalence in MatchKAT is $\mathsf{PSPACE}$-complete.\end{abstract}

\section{Introduction}

The \emph{match-action} paradigm has remained a popular low-level
programming model for specifying packet forwarding behavior in network
switches. In this model, a switch is organized as one or more \emph{match
tables} in sequence, each containing rules with \emph{patterns} and
\emph{actions}. The pattern is some match specification on the binary
data fields in a packet header, such as a ternary expression containing
$0$, $1$ or don't-care. The action is some modification on the packet
header. When a packet arrives at a match table, a rule is selected
among those with matching patterns and the associated action is executed.
The selection criterion could be some pre-configured \emph{priority}
ordering on the rules, or based on some property of the pattern such
as selecting the one with the fewest don't-cares (longest prefix matching)
\cite{McKeown:2008:OEI:1355734.1355746}.

There are efficient hardware implementations of match-action \cite{Lakshminarayanan:2005:AAP:1080091.1080115},
and it is a simple model accepted by network programmers. Nevertheless,
high-level domain specific languages (DSLs) such as NetKAT \cite{Anderson:2014:NSF:2535838.2535862}
and P4 \cite{Bosshart:2014:PPP:2656877.2656890} are available to
provide abstractions for network policies that can then be compiled
down to match-action tables in the target switch \cite{Choi:2017:PPV:3050220.3060609,Shahbaz:2016:PPP:2934872.2934886,Smolka:2015:FCN:2784731.2784761}.
Despite much theoretic work surrounding these DSLs, there has been
comparatively little investigation towards putting match-action itself
on a firm theoretical foundation.

Towards the goal of formalizing match-action, we present MatchKAT,
a Kleene algebra with tests (KAT) that employs match expressions on
binary strings as tests. It is able to encode match and action while
having a metatheory closely related to NetKAT. Leveraging results
from NetKAT, we are able to show MatchKAT is sound and complete with
respect to its own packet filtering semantics. Through a translation
to NetKAT, decision procedures such as those in \cite{Foster:2015:CDP:2676726.2677011}
can also be adapted to MatchKAT. Although this paper will mainly introduce
the basics of MatchKAT and its metatheory, the application-level motivation
is that in the future we may be able to give a formal semantics for
match-action as used in network switches. It is hoped that MatchKAT
will eventually allow for algebraic reasoning on local switch configurations
similar to NetKAT for global network policies, which could allow applications
such as proving the equivalence of match-action switch configurations
and decompiling match-action rules to higher-level policies. Previous
attempts at reasoning with match expressions on binary strings in
the context of packet classification, such as in \cite{KazemianPhD,180587},
have been more ad hoc and without a formal metatheory.

Our contributions can be summarized as follows:
\begin{itemize}
\item We give an algebraic formalization of ternary ($0$, $1$, don't-care)
match expressions on binary strings (Section \ref{sec:Match-Expressions}).
Although others have studied aspects of the theory of match expressions,
for example \cite{180587}, we present it here in a formal algebraic
language as match expressions will be integral to the formalization
of MatchKAT.
\item We give the syntax and a packet filtering semantics for MatchKAT (Section
\ref{sec:MatchKAT}), and show how it is able to to encode match-action
(Section \ref{subsec:Encoding-Tables-and}). Despite being related
to NetKAT, MatchKAT is able to encode operations that would require
much longer expressions in NetKAT.
\item We show that MatchKAT has a sound and complete equational theory with
respect to its semantics by leveraging a correspondence with the $\mathsf{dup}$-free
fragment of NetKAT (Sections \ref{sec:Connection-with-NetKAT} and
\ref{sec:Equational-Theories-of}). The problem of deciding equivalence
between MatchKAT terms is shown to be $\mathsf{PSPACE}$-complete
(Section \ref{subsec:Complexity-of-Deciding}).
\end{itemize}

\section{Preliminaries}

In this section we give some background on KATs, as well as a formal
presentation of match expressions on binary strings. We will defer
discussion on NetKAT to Section \ref{sec:Connection-with-NetKAT}
when we clarify its connection with MatchKAT.

\subsection{Kleene Algebras with Tests}

A Kleene algebra with tests (KAT) \cite{Kozen:1997:KAT:256167.256195}
has a signature $\left(P,B,+,\cdot,^{*},\boldsymbol{0},\boldsymbol{1},\barnone\right)$
such that
\begin{itemize}
\item $\left(P,+,\cdot,^{*},\boldsymbol{0},\boldsymbol{1}\right)$ is a
Kleene algebra.
\item $\left(B,+,\cdot,\barnone,\boldsymbol{0},\boldsymbol{1}\right)$ is
a Boolean algebra.
\item $\left(B,+,\cdot,\boldsymbol{0},\boldsymbol{1}\right)$ is a subalgebra
of $\left(P,+,\cdot,\boldsymbol{0},\boldsymbol{1}\right)$.
\end{itemize}
$P$ is usually called the set of \emph{primitive actions }while members
of $B$ are \emph{primitive tests. }Note that $\boldsymbol{0}$ and
$\boldsymbol{1}$ are the identities of $+$ and $\cdot$ respectively
and $\boldsymbol{0}$ is an annihilator for $\cdot$. Terms of the
KAT are then freely generated by $P$ and $B$ with the operators.
We omit most of the algebraic theories here as they are well-covered
in literature \cite{conway1971regular,Kozen:1997:KAT:256167.256195,Kozen:1997:AC:549365}.
We will however highlight that KATs can possess interesting equational
theories, as we will be studying later. It is possible to axiomatically
derive equivalences between KAT terms, as well as assign some denotational
semantics to them. We say that the equational theory is \emph{sound}
with respect to those semantics if all provably equal terms have equal
semantics, and \emph{complete} if proofs of equivalence exist for
any two terms that are semantically equal. The decision problem of
whether two terms are equal can also be studied and its complexity
classified. Since KATs can often be used to encode programs, the equational
theory is important for studying program equivalence.

\subsection{Match Expressions\label{sec:Match-Expressions}}

We give a formalization of match expressions on binary strings as
found in match-action tables implemented in network switches. These
expressions will form the tests within MatchKAT.

The set $\mathbb{E}$ will be the set of all match expressions that
we will define. The syntax of expressions is found in Figure \ref{fig:Syntax-of-match}.
$\mathbb{E}$ is equipped with a concatenation operation $@$ and
is stratified into subsets $E_{n}$ for all $n\geq0$, such that $\mathbb{E}\triangleq\bigcup_{n}E_{n}.$
Each $E_{n}$ is said to be \emph{the set of match expressions with
width $n$}, and has an algebraic signature $\left(E_{n},+,\sqcap,\barnone,\bot,\top_{n}\right)$.
$+$ is\emph{ union}, $\sqcap$ is \emph{intersection}, $\barnone$
is \emph{complementation}, and $\bot$ and $\top_{n}$ are identities
of $+$ and $\sqcap$ respectively. Terminology-wise, we will refer
to the size of binary strings to be matched on as the \emph{width},
reserving the word \emph{length} for later quantifying the size of
match expressions themselves.

\begin{figure}
\[
E_{0}::=\text{\ensuremath{\bullet}}\mid\bot\mid E_{0}+E_{0}\mid E_{0}\sqcap E_{0}\mid\bar{E}_{0}
\]
\[
E_{n+1}::=E_{n}\,@\,1\mid E_{n}\,@\,0\mid E_{n}\,@\,\mathrm{x}\mid\bot\mid E_{n+1}+E_{n+1}\mid E_{n+1}\sqcap E_{n+1}\mid\bar{E}_{n+1}
\]
\[
\begin{array}{ccc}
\top_{0}\triangleq\bullet & \,\,\,\,\,\,\,\,\,\,\, & \top_{n}\triangleq\underset{n}{\underbrace{\mathrm{x}\dots\mathrm{x}}}\text{ for all }n>0\end{array}
\]
\caption{\label{fig:Syntax-of-match}Syntax of match expressions.}
\end{figure}

The actual members of the set $E_{n}$ are defined inductively on
the width $n$. In the base case, there are the empty $\bullet$ and
bottom $\bot$ expressions. Note that we distinguish between the empty
expression and the empty binary string $\epsilon$. $E_{n+1}$ is
then built from members of $E_{n}$ with concatenation $@$. Notationally
we will usually elide this operator.

Intuitively, $1$, $0$ and $\mathrm{x}$ will correspond to matching
$1$, $0$ or anything (don't-care) at a given position in the binary
string, $\bullet$ is for matching $\epsilon$, and $\bot$ matches
nothing. We use $\text{x}$ to avoid confusion with the $^{*}$ operator
of KATs. This intuition of an expression matching bits will be made
formal shortly. Since $E_{n}$ at each width has the signature $\left(E_{n},+,\sqcap,\barnone,\bot,\top_{n}\right)$,
it is also extended freely with expressions built from $+$, $\sqcap$
and $\barnone$.

Axiomatically, for every $n$ we require $\left(E_{n},+,\sqcap,\barnone,\bot,\top_{n}\right)$
to be a Boolean algebra. The Boolean algebra axioms determine the
behavior of $\bot$ and $\top_{n}$ when combined with the Boolean
operators. However, additionally we also need axioms that relate $1$,
$0$, $\mathrm{x}$ and concatenation. They are found in Figure \ref{fig:Axioms-of-match}.
Note that axiomatically all expressions concatenated with $\bot$
collapse to just $\bot$. It is therefore unnecessary to distinguish
$\bot$s of different widths. On the other hand $\top_{n}$ is syntactic
sugar for the wildcard expression matching any string of width $n$,
and there is a distinct such expression for each $n$.
\begin{figure}
\[
\begin{array}{lllll}
\bar{0}=1 & \,\,\,\,\,\,\,\,\,\, & 1+0=0+1=\mathrm{x} & \,\,\,\,\,\,\,\,\,\, & 1\sqcap0=0\sqcap1=\bot\end{array}
\]
\[
\begin{array}{ccccccc}
\bullet e=e\bullet=e & \,\,\,\,\,\,\,\,\,\, & \bot e=e\bot=\bot & \,\,\,\,\,\,\,\,\,\, & \overline{ee'}=\bar{e}\top_{n_{2}}+\top_{n_{1}}\bar{e'} & \,\,\,\,\,\,\,\,\,\, & \left(e_{1}e_{2}\right)e_{3}=e_{1}\left(e_{2}e_{3}\right)\end{array}
\]
\[
\begin{array}{ccc}
e_{1}\left(e_{2}+e_{3}\right)=e_{1}e_{2}+e_{1}e_{3} & \,\,\,\,\,\,\,\,\,\,\, & e_{1}\left(e_{2}\sqcap e_{3}\right)=e_{1}e_{2}\sqcap e_{1}e_{3}\\
\left(e_{1}+e_{2}\right)e_{3}=e_{1}e_{3}+e_{2}e_{3} &  & \left(e_{1}\sqcap e_{2}\right)e_{3}=e_{1}e_{3}\sqcap e_{2}e_{3}
\end{array}
\]
\caption{\label{fig:Axioms-of-match}Axioms for match expressions for $e_{1}$,
$e_{2}$, $e_{3}$ in $\mathbb{E}$, $e$ in $E_{n_{1}}$, and $e'$
in $E_{n_{2}}$. These are in addition to axioms that enforce $\left(E_{n},+,\sqcap,\barnone,\bot,\top_{n}\right)$
for each $n$ as Boolean algebras.}
\end{figure}

\noindent \begin{flushleft}
To formalize the semantics of match expressions, we come back to the
notion of width and length as mentioned at the start of this section.
Let $2^{n}$ be the set of binary strings of width $n$, and in particular
let $2^{0}=\left\{ \epsilon\right\} $ be the set only containing
the empty string $\epsilon$. An expression $e\in E_{n}$ is said
to have width $n$ and matches strings in $2^{n}$.
\par\end{flushleft}

We can model what it means for a match expression to match a binary
string by interpreting an expression as the set of all strings that
match it. For some expression $e\in E_{n}$, its interpretation $\llp e\rrp$
is the set of all strings in $2^{n}$ that matches $e$. The definition
of $\llp e\rrp\subseteq2^{n}$ is made inductively on $e$:
\[
\begin{array}{rclcrclcrcl}
\llp\bullet\rrp & \triangleq & \left\{ \epsilon\right\}  & \,\,\,\,\,\,\,\,\,\,\, & \llp\text{x}\rrp & \triangleq & \left\{ 0,1\right\}  & \,\,\,\,\,\,\,\,\,\,\, & \llp e+e'\rrp & \triangleq & \llp e\rrp\cup\llp e'\rrp\\
\llp0\rrp & \triangleq & \left\{ 0\right\}  &  & \llp\bot\rrp & \triangleq & \emptyset &  & \llp e\sqcap e'\rrp & \triangleq & \llp e\rrp\cap\llp e'\rrp\\
\llp1\rrp & \triangleq & \left\{ 1\right\}  &  & \llp ee'\rrp & \triangleq & \left\{ xx'\mid x\in\llp e\rrp,x'\in\llp e'\rrp\right\}  &  & \llp\bar{e}\rrp & \triangleq & 2^{n}-\llp e\rrp
\end{array}
\]

\begin{example}
Since $\top_{n}\triangleq\underset{n}{\underbrace{\mathrm{x}\dots\mathrm{x}}}\text{ for }n>0$,
$\llp\top_{n}\rrp=2^{n}$ by derivation from the definitions $\llp\text{x}\rrp$
and $\llp ee'\rrp$.
\end{example}
For any $S\subseteq2^{n}$, we say that $e$ \emph{captures }$S$
if and only if $\llp e\rrp=S$. When reasoning with binary strings,
we often wish to refer to individual bits within the string. For $b\in2^{n}$,
we write $b\left[i\right]$ for the $i$-th bit of $b$, and $b\left[i\leftarrow x\right]$
for the string that is $b$ but with $x$ as the $i$-th bit. Conventionally,
we will use base-1 for bit indices, strings are read left-to-right,
and the most significant bit is to the left whenever a string is interpreted
as a binary number.

Therefore it can be seen that a match expression has the same width
as the strings it matches. On the other hand, its length could be
arbitrary in size, and it is a measure of the complexity of the expression.
\begin{example}
Suppose we are interested only in counting occurrences of $\sqcap$
and $+$. For some even width $2n$, consider the expression
\[
\bigsqcap_{i=1}^{n}\left(\top_{i-1}0\top_{n-1}0\top_{n-i}+\top_{i-1}1\top_{n-1}1\top_{n-i}\right).
\]
It captures exactly the set $\left\{ bb\mid b\in2^{2n}\right\} $
and its length is $O\left(n\right)$ since that many $\sqcap$ and
+ operators were used. An equivalent expression that captures the
same set is
\[
\overline{\sum_{i=1}^{n}\left(\top_{i-1}1\top_{n-1}0\top_{n-i}+\top_{i-1}0\top_{n-1}1\top_{n-i}\right)},
\]
which is also length $O\left(n\right)$. However, if we are only allowed
to use $+$, but not $\sqcap$ and complementation, an expression
capturing this set must have length at least exponential in $n$.
This is because each string of the form $bb$ must occur in the match
expression explicitly.

The same match expression could have different lengths depending on
which operators we are interested in counting. This is useful for
the application of relating match expression length to the complexity
of a match program in a network switch. Some operations may be expensive,
such as $\sqcap$, while concatenation can be ``free'' and do not
need to be counted as it is simply multiple hardware units placed
in parallel.
\end{example}
We end the discussion on match expressions by speaking briefly on
the soundness and completeness of the equational theory of match expressions
with respect to the binary strings model. Proving soundness is a straightforward
albeit tedious task. We simply go through each axiom and show that
the expressions on both sides of the equality capture the same set.
Completeness is also fairly easy. We can decide whether two match
expressions $e$ and $e'$ are equivalent by expanding both to their
disjunctive normal forms and then eliminate all occurrences of $\sqcap$.
Equality can then be checked if the expressions are identical up to
commutativity of $+$. Unfortunately, this axiomatic proof of equivalent
introduces an exponential blowup. A more tractable, co-NP decision
procedure is to non-deterministically guess a string in the symmetric
difference of $\llp e\rrp$ and $\llp e'\rrp$, which succeeds if
and only if $e$ and $e'$ are not equivalent.

\section{MatchKAT\label{sec:MatchKAT}}

Our discussion of MatchKAT starts with the intuition that each width-$n$
space of match expressions $E_{n}$ can be seen as a Boolean algebra
over $n$ variables. A binary string corresponds to an assignment
of truth values and a match expression is a propositional formula
that is satisfied by exactly the assignments of matching binary strings.
This is an alternative way to think of the underlying model that we
are working with as our definition of MatchKAT evolves.

\subsection{Definitions\label{subsec:matchkat-Definitions}}

Let $n$ be a constant positive integer, which as before was used
to denote the widths of binary strings, but now we will refer to it
as the \emph{packet size}. Intuitively, MatchKAT is a KAT whose terms
operate on the finite state space created by $n$ bits of random access
memory occupied by a packet header. It is defined by:
\begin{itemize}
\item Primitive tests are match expressions in $E_{n}$, matching the whole
memory at once. For $1\leq i\leq n$ and $k\in\left\{ 0,1\right\} $,
we will adopt the shorthand $i\simeq k$ for the match expression
$\top_{i-1}k\top_{n-i}$, which solely tests whether the $i$-th bit
is $k$.
\item Primitive actions are in the form $i\leftarrow k$, for $1\leq i\leq n$
and $k\in\left\{ 0,1\right\} $, intended to mean assigning $0$ or
$1$ to bit $i$.
\item The operations are plus $+$, composition $\cdot$, complementation
$\bar{p}$, and Kleene star $p^{*}$. For tests, $+$ and $\cdot$
correspond respectively to $+$ and $\sqcap$ within match expressions
$E_{n}$ (not concatenation within $\mathbb{E}$). Sometimes we may
write composition as $\sqcap$ between terms that are known to be
tests.
\item The identity of $+$ is $\bot$, and for $\cdot$ it is $\top_{n}$,
or just $\top$ for short.
\end{itemize}
We admit all the axioms required of a KAT, and those of match expressions
presented previously. This is already a sufficient definition for
a valid KAT. However, we require additional \emph{packet algebra axioms}
in order to allow commutation of actions and tests on unrelated memory
locations, and absorption of related ones. For $i\neq j$:
\[
\begin{array}{rclcrcl}
i\leftarrow k\cdot j\leftarrow k' & \equiv & j\leftarrow k'\cdot i\leftarrow k & \,\,\,\,\,\,\,\,\,\, & i\leftarrow k\cdot j\simeq k' & \equiv & j\simeq k'\cdot i\leftarrow k\\
i\leftarrow k\cdot i\simeq k & \equiv & i\leftarrow k &  & i\simeq k\cdot i\leftarrow k & \equiv & i\simeq k
\end{array}
\]

We use $\equiv$ to denote the equivalence of terms in order to avoid
ambiguity with $\simeq$ and $=$. Readers familiar with NetKAT may
wonder why we do not require axioms of the forms $i\simeq k\cdot i\simeq k\equiv i\simeq k$,
$k\neq k'\implies i\simeq k\cdot i\simeq k'\equiv\bot$, and $\sum_{k}i\simeq k\equiv\top$.
These are derivable theorems within the algebra of match expressions.
\begin{example}
\label{exa:NetKAT-axiom-thm}If $k\neq k'$, then 
\[
i\simeq k\cdot i\simeq k'\equiv\left(\top_{i-1}k\top_{n-i}\right)\sqcap\left(\top_{i-1}k'\top_{n-i}\right)\equiv\top_{i-1}\left(k\sqcap k'\right)\top_{n-i}\equiv\bot
\]
 as $k\sqcap k'\equiv\bot$.
\end{example}

\subsection{Packet Filtering Semantics}

We now discuss the semantics of MatchKAT as applied to packet forwarding.
Naturally, the $n$ bits of state we have in mind will be modeled
by packet headers, which we will just refer to as packets. The following
semantics operate on sets of packets at both input and output, intending
to model the packets that arrive at a switch and what packets will
be forwarded after filtering by the MatchKAT term. We denote the set
of packets as $Pk$, which we will represent as strings in $2^{n}$
(so really $Pk=2^{n}$). The semantics of a MatchKAT term $e$ is
a function $\llb e\rrb:\mathcal{P}\left(Pk\right)\to\mathcal{P}\left(Pk\right)$:
\[
\begin{array}{rclcrcl}
\llb\bot\rrb\left(P\right) & \triangleq & \emptyset & \,\,\,\,\,\,\,\,\,\,\, & \llb p+q\rrb\left(P\right) & \triangleq & \llb p\rrb\left(P\right)\cup\llb q\rrb\left(P\right)\\
\llb\top\rrb\left(P\right) & \triangleq & P &  & \llb p\cdot q\rrb\left(P\right) & \triangleq & \left(\llb q\rrb\circ\llb p\rrb\right)\left(P\right)\\
\llb a\rrb\left(P\right) & \triangleq & P\cap\llp a\rrp,\,a\in E_{n} &  & \llb\bar{p}\rrb\left(P\right) & \triangleq & Pk-\llb p\rrb\left(P\right)\\
\llb i\leftarrow k\rrb\left(P\right) & \triangleq & \left\{ \pi\left[i\leftarrow k\right]\mid\pi\in P\right\}  &  & \llb p^{*}\rrb\left(P\right) & \triangleq & \bigcup_{k\geq0}\llb p\rrb^{k}\left(P\right)
\end{array}
\]
We call this the \emph{packet filtering semantics} as the semantic
functions are transformers on sets of packets. Suppose a network switch
is modeled by a MatchKAT term, the output denotes the set of packets
that is produced given some set of input packets. The next sections
will give examples of how MatchKAT terms can be used in practice,
while later in Section \ref{sec:Equational-Theories-of} we will show
the equational theory of MatchKAT is sound and complete with respect
to this semantics, and deciding equivalence is $\mathsf{PSPACE}$-complete.

\subsection{Encoding Actions on Packets}

In match-action, ``match'' refers to matching of binary data in
packet headers, which we have covered so far. On the other hand, ``actions''
in this context refer to simple modifications of the packet header.
Once a rule is matched, its action is performed and the switch then
forwards (or keeps on processing) the packet based on the updated
header fields. For example, there may be a \emph{port} field specifying
the egress port the packet should be moved. It is possible to encode
modifications on fields in MatchKAT.

\subsubsection{Direct and indirect assignment/test.}

Assignment/test of a constant value over a range of bits can be performed
by assigning/testing the value's binary representation.
\begin{example}
Assigning the value 6 (binary $110$) to bits $2$ through $4$ can
be written as $2\leftarrow1\cdot3\leftarrow1\cdot4\leftarrow0$.
\end{example}
Test and assignment of a range of bits against another range can be
done in a single match expression bitwise.
\begin{example}
To assign the values contained in bits $1$ through $3$ to bits $4$
through $6$, we can write

\[
\begin{array}{ll}
 & \left(1\simeq0\cdot4\leftarrow0+1\simeq1\cdot4\leftarrow1\right)\\
\cdot & \left(2\simeq0\cdot5\leftarrow0+2\simeq1\cdot5\leftarrow1\right)\\
\cdot & \left(3\simeq0\cdot6\leftarrow0+3\simeq1\cdot6\leftarrow1\right).
\end{array}
\]
We can simply replace $\leftarrow$ with $\simeq$ above instead to
test for equality.
\end{example}

\subsubsection{Arithmetic.}

Since we know in advance the packet size $n$, and the range of bits
to operate on, we can encode arithmetic on sets of bits in MatchKAT
through simple fixed-width algorithms. We give incrementation just
as an example.
\begin{example}
Suppose a range of bits contains a binary value we wish to increment.
We write $\left[i\dots j\right]^{++}$ for the term that increments
the value contained in bits $i$ through $j$. It can be defined inductively
as:
\[
\left[i\dots j\right]^{++}\triangleq\begin{cases}
\begin{array}{l}
\top\\
j\simeq0\cdot j\leftarrow1+j\simeq1\cdot j\leftarrow0\cdot\left[i\dots j-1\right]^{++}
\end{array} & \begin{array}{r}
j<i\\
\text{otherwise}
\end{array}\end{cases}
\]
\end{example}

\subsection{Encoding Match-Action Tables\label{subsec:Encoding-Tables-and}}

In real match-action tables, match patterns and actions are paired
in rules. A single rule can be easily encoded in MatchKAT as the composition
of a test with actions. Less straightforward is capturing the rule
selection mechanism of the table. For example, let match expressions
be $b_{1}\dots b_{k}$ and actions $p_{1}\dots p_{k}$. In a table
with rules $\left(b_{1}p_{1}\right)\dots\left(b_{k}p_{k}\right)$,
we may have multiple $b$ expressions matching an incoming packet.
In a \emph{priority}-ordered table, the rule that is actually selected
and has its action executed is based on some pre-assigned priority
ordering on the rules. Here suppose $1$ is the highest priority and
$k$ the lowest. A naive MatchKAT encoding of the table as $b_{1}p_{1}+\dots+b_{k}p_{k}$
does not work, since in a KAT $+$ is commutative. To impose an order,
the simplest way is to negate all higher-priority tests:
\[
b_{1}p_{1}+\overline{b_{1}}b_{2}p_{2}+\dots+\overline{b_{1}}\dots\overline{b_{k-1}}b_{k}p_{k}.
\]
This term contains $O\left(k\right)$ sums and $O\left(k^{2}\right)$
compositions. Albeit inefficient, in this case indeed a rule's action
will only be executed if no higher-priority rule matched.

An alternative encoding is to set aside some \emph{metadata} bits
as a counter to record the current rule being matched. Suppose this
counter resides in bits $i$ through $j$, then using incrementation
from the previous section, we can write:
\[
\left(\left[i\dots j\right]\leftarrow1\right)\left[\sum_{r=1}^{k}\left(\left[i\dots j\right]=r\cdot\left(b_{r}p_{r}\cdot\left[i\dots j\right]\leftarrow\left(k+1\right)+\overline{b_{1}}\left[i\dots j\right]^{++}\right)\right)\right]^{*}.
\]
Here we write $\left[i\dots j\right]=r$ as shorthand for testing
the range bitwise for the binary number $r$. The encoded term works
by only testing rule $r$ if $\left[i\dots j\right]$ has value $r$.
If $b_{r}$ succeeds then action $p_{r}$ is executed, and the rule
counter is set to the end value $k+1$. If $b_{r}$ fails then the
rule counter is incremented. Kleene star is used to iterate through
all the rules.

The above examples are not the only possible ways to encode match-action
tables in MatchKAT. However, since we will prove that the equational
theory of MatchKAT is sound and complete with respect to its packet
filtering semantics, in principle we should be able to prove equivalence
between all possible valid encodings. Even though different encodings
have equivalent semantics, they may have different implementation
qualities such as the length of match expressions, the depth of nesting,
and the use of additional bits to store metadata such as $\left[i\dots j\right]$
in the example above. Nevertheless we can establish a notion of program
equivalence between these two ways of representing a table of match-action
rules.

In some network switches there exist more than one match-action table
organized in a \emph{pipeline} \cite{Bosshart:2014:PPP:2656877.2656890,McKeown:2008:OEI:1355734.1355746}\emph{.}
The tables can be sequentially composed, or possibly be in parallel
with branching and loops. These can all be handled by MatchKAT's $\cdot$
for sequencing, $+$ for parallelism or branching, and $^{*}$ for
loops.

\section{Connection with NetKAT\label{sec:Connection-with-NetKAT}}

NetKAT is an algebraic language based on Kleene algebra with tests
that is able to specify packet forwarding policies in a network \cite{Anderson:2014:NSF:2535838.2535862}.
Before we study the equational theory of MatchKAT, we will precisely
define a connection between MatchKAT and NetKAT in both a syntactic
and also semantic sense. This will allow us to leverage known results
about NetKAT in the MatchKAT setting. Syntactically, there is a correspondence
between MatchKAT and the $\mathsf{dup}$-free fragment of NetKAT,
and we will elaborate on this shortly. Semantically, NetKAT is mainly
concerned with the possible progressions of a packet through the network,
whereas we are more interested in the behavior of a single, local
switch on packets. The syntactic and semantic relationships are entirely
consistent. $\mathsf{dup}$ can be used in NetKAT to record the states
of a packet at different hops, so it is natural that without $\mathsf{dup}$,
we instead reason about what happens on the local hop. This is referred
to in \cite{Smolka:2015:FCN:2784731.2784761} as the ``local program'',
where the switch configuration is still in NetKAT but agnostic about
the network topology. However, we emphasize that MatchKAT is not intended
to serve the same purpose as NetKAT. The language instead focuses
on lower-level match expressions and manipulation of bits as this
is closer to what is implemented in hardware.

We give a short description of NetKAT's syntax and its axioms, but
since NetKAT is well-presented elsewhere, we will not discuss too
many details here. What we will see at by the end of this section,
however, are mutual translations between MatchKAT and NetKAT that
will come in useful when we study MatchKAT's equational theory.

\subsection{Syntax and Axioms of NetKAT}

Let $F=\left\{ f_{1},f_{2},\dots,f_{n}\right\} $ be some fixed, finite
set of \emph{fields}. NetKAT is a KAT again with signature $\left(P,B,+,\cdot,^{*},\boldsymbol{0},\boldsymbol{1},\barnone\right)$
whose primitive tests $B$ and actions $P$ are defined with respect
to $F$:
\begin{itemize}
\item In addition to $\boldsymbol{0}$ and $\boldsymbol{1}$, primitive
tests are of the form $f_{i}=k$, for some natural number $k$ and
$f_{i}\in F$.
\item There is a special primitive action named $\mathsf{dup}$. Other primitive
actions are in ``assignment'' form $f_{i}\leftarrow k$.
\end{itemize}
We assume for each field $f_{i}$ there exists a finite set of natural
numbers that could be associated with the field. Hence a NetKAT term
is not well-formed if it contains $f_{i}=k$ or $f_{i}\leftarrow k$
for $k$ not in that set. Just like MatchKAT, in addition to the standard
KAT axioms, NetKAT requires packet algebra axioms governing mainly
when tests and actions can commute. They can be found in \cite{Anderson:2014:NSF:2535838.2535862}
and it suffices for us to say that they are similar to those in MatchKAT,
except for one additional axiom involving $\mathsf{dup}$.

We highlight the fact that tests and actions in NetKAT involve constant
values. At first glance this may appear more limited than MatchKAT's
ability to perform indirect assignment and computation on fields as
demonstrated previously. We point out that this is only possible in
MatchKAT's case since the size $n$ of the state space is known and
we are performing fixed-width arithmetic. Although we will see later
that there is a close connection between the two, this difference
in focus between NetKAT and MatchKAT means they are still separate
languages dealing with different levels of abstraction of network
programs.

\subsection{Semantics}

We will talk briefly about the semantics of NetKAT, while readers
interested in a formal detailed treatment are invited to read \cite{Anderson:2014:NSF:2535838.2535862}.
In NetKAT, a \emph{packet} is a record of field-value pairs $\left\{ f_{1}=k_{1},\dots,f_{n}=k_{n}\right\} $
where each field has a valid assignment of values. This represents
the header of a real-life packet that is of interest when we are deciding
on its forwarding behavior. A \emph{packet history} is simply a list
of packets with the head being the most recent.
\begin{defn}
Let $\mathsf{H}$ be the set of packet histories. For $\pi\in Pk$,
we write $\pi::\left\langle \right\rangle $ for the packet history
with $\pi$ at its head and nothing else, and $hd$ to be the function
that takes packet history to their head packets. When we conflate
notation and write $hd\,H$ for $H\subseteq\mathsf{H}$, we mean the
set $\left\{ hd\,h\mid h\in H\right\} $.
\end{defn}
In NetKAT's packet filtering semantics, the interpretation of a term
$e$ is a function $\llb e\rrb:\mathsf{H}\to\mathcal{P}\left(\mathsf{H}\right)$.
Composition of these functions is done through Kleisli composition
in the powerset monad. The semantics can be thought of as the behavior
of a switch when it is presented with the head packet in a packet
history. Each packet in the history represents a previous state of
the head packet, possibly at a previous switch in the network. Using
packet histories, as opposed to simply packets, allows us to distinguish
packets that have taken different paths in the network. However, the
input history beyond the head packet cannot be accessed directly by
NetKAT terms, consistent with a switch not being able to see the operations
that previous switches have done to the packet.

The semantics of NetKAT can be explained intuitively. $\boldsymbol{1}$
lets a packet through unchanged, while $\boldsymbol{0}$ drops the
packet. $f=k$ and $f\leftarrow k$ tests and assigns the field $f$
with the value $k$ respectively, in the head packet of the input
packet history. $\mathsf{dup}$ duplicates the current head packet
and places a copy of it at the head of the history, i.e.
\[
\llb\mathsf{dup}\rrb\left(\pi::h\right)\triangleq\left\{ \pi::\pi::h\right\} .
\]
Note also that the codomain of the semantic function is sets of packet
histories. This accommodates the fact that it is possible for a switch
to egress multiple packets in response to a packet at ingress, possibly
different in content and to different destinations. Composition $\cdot$
of interpretations having type $\mathsf{H}\to\mathcal{P}\left(\mathsf{H}\right)$
is done through Kleisli composition in the powerset monad, in contrast
to function composition in MatchKAT. $+$ and $\barnone$ becomes
union and complementation in the result sets respectively, and $^{*}$
takes the usual meaning of iterated composition.
\begin{example}
Suppose the set of fields is $\left\{ \mathsf{pt},\mathsf{proto},\mathsf{ttl}\right\} $
and $\mathsf{pt}$ is understood to be the switch port where the packet
is located. The NetKAT term $\mathsf{pt}=1\cdot\mathsf{proto}=6\cdot\mathsf{dup}\cdot\mathsf{ttl}\leftarrow40\cdot\mathsf{pt}=3$
is the policy ``If the packet is at port $1$ and has $\mathsf{proto}$
value 6, take a snapshot of its current state, change the $\mathsf{ttl}$
value to 40 and move the packet to port $3$. Otherwise drop the packet.''
\end{example}

\subsection{MatchKAT to NetKAT}

We will now formally define a translation from MatchKAT to NetKAT.
For a MatchKAT with packet size $n$, the corresponding NetKAT will
be over $n$ fields, $f_{1}$ through $f_{n}$, each taking $0$ or
$1$ in value. We define a homomorphism $\lceil\cdot\rceil$ that
takes terms in this MatchKAT to the corresponding NetKAT terms as
follows:
\[
\begin{array}{ccccc}
\lceil\bot\rceil\triangleq\boldsymbol{0} & \,\,\,\,\,\,\,\,\,\, & \lceil\top\rceil\triangleq\boldsymbol{1} & \,\,\,\,\,\,\,\,\,\, & \lceil i\leftarrow k\rceil\triangleq f_{i}\leftarrow k\end{array}
\]
The definitions for $+$, $\cdot$, $^{*}$ and $\barnone$ terms
extend homomorphically, i.e.
\[
\begin{array}{ccccccc}
\lceil e+e'\rceil\triangleq\lceil e\rceil+\lceil e'\rceil & \,\,\,\,\,\,\,\,\,\, & \lceil e\cdot e'\rceil\triangleq\lceil e\rceil\cdot\lceil e'\rceil & \,\,\,\,\,\,\,\,\,\, & \lceil e^{*}\rceil\triangleq\lceil e\rceil^{*} & \,\,\,\,\,\,\,\,\,\, & \lceil\overline{e}\rceil\triangleq\overline{\lceil e\rceil}\end{array}.
\]
 We complete the definition for primitive tests by giving the translation
in terms of match expressions on single bits and then concatenation.
Translations of more complex match expressions extend naturally from
the definitions for $+$ and $\cdot$.
\[
\begin{array}{ccccccc}
\lceil0\rceil\triangleq f_{i}=0 & \,\,\,\,\,\,\,\,\,\, & \lceil1\rceil\triangleq f_{i}=1 & \,\,\,\,\,\,\,\,\,\, & \lceil\text{x}\rceil\triangleq\boldsymbol{1} & \,\,\,\,\,\,\,\,\,\, & \lceil e@e'\rceil\triangleq\lceil e\rceil\cdot\lceil e'\rceil\end{array}
\]
Here $i$ refers to the bit position that the single-bit expression
$0$ or $1$ is matching. We can pre-compute these position values
for every $0$ or $1$ that appears in the expression before carrying
out the translation. Notice that the translation does not introduce
any $\mathsf{dup}$s, and is a straightforward syntactic embedding
into NetKAT. More importantly, this translation is semantic preserving
in the following way.
\begin{thm}
\label{thm:matchkat-to-netkat}For any MatchKAT term $e$, $\llb e\rrb\left(P\right)=\bigcup_{\pi\in P}hd\left[\llb\lceil e\rceil\rrb\left(\pi::\left\langle \right\rangle \right)\right]$.
\end{thm}
The proof is a standard induction on $e$. We will simply observe
that since the translation introduces no $\mathsf{dup}$s, $hd\left(\pi::\left\langle \right\rangle \right)=\pi$,
and it is clear that $\lceil e\rceil$ performs the same operations
in NetKAT as $e$ does in MatchKAT.

\subsection{NetKAT to MatchKAT}

Similarly, there is a translation from NetKAT to MatchKAT. Since the
latter is $\mathsf{dup}$-free, such a translation is forgetful in
the sense that we lose the packet history structure entirely and only
track the state of the head packet.

Suppose the particular NetKAT we wish to translate from has fields
$f_{1}$ through $f_{m}$. We assume it is possible to represent the
values in each field in binary, and let $\left|f_{i}\right|$ denote
the number of bits required to store $f_{i}$. We set the target MatchKAT
packet size to be $n=\sum_{i}\left|f_{i}\right|$. The translation
from NetKAT terms to MatchKAT terms is again a homomorphic function
$\lfloor\cdot\rfloor$, and it is only necessary for us to specify
its action on the on primitives:
\[
\begin{array}{ccccc}
\lfloor\boldsymbol{0}\rfloor\triangleq\bot & \,\,\,\,\,\,\,\,\,\, & \lfloor\boldsymbol{1}\rfloor\triangleq\top & \,\,\,\,\,\,\,\,\,\, & \lfloor\mathsf{dup}\rfloor\triangleq\top\end{array}
\]
\[
\begin{array}{ccc}
\lfloor f_{i}=k\rfloor\triangleq\bigsqcap_{j=1}^{\left|f_{i}\right|}pos_{j}\left(f_{i}\right)\simeq bin_{j}\left(k\right) & \,\,\,\,\,\,\,\,\,\, & \lfloor f_{i}\leftarrow k\rfloor\triangleq\prod_{j=1}^{\left|f_{i}\right|}pos_{j}\left(f_{i}\right)\leftarrow bin_{j}\left(k\right)\end{array}
\]
The function $pos_{j}\left(f_{i}\right)$ gives bit position for the
$j$-th bit in the header space allocated for $f_{i}$, i.e. it is
$pos_{j}\left(f_{i}\right)=j+\sum_{i'<i}\left|f_{i'}\right|$.

On the other hand, $bin_{j}\left(k\right)$ is the $j$-th bit of
the binary representation of $k$. The translation for assignment
$f_{i}\leftarrow k$ just sets each bit in the space allocated for
$f_{i}$ in the target MatchKAT bitwise. This is the same method for
test $f_{i}=k$, with the resulting bitwise tests composed by $\sqcap$,
and we can always equivalently combine the tests into a match expression
without $\sqcap$ by using match expression axioms like in Example
\ref{exa:NetKAT-axiom-thm}. We forget the existence of $\mathsf{dup}$
by translating it as $\top$. Just like the translation to NetKAT,
$\lfloor\cdot\rfloor$ implies a semantic correspondence.
\begin{thm}
\label{thm:netkat-to-matchkat}For any NetKAT term $e$, $hd\left[\llb e\rrb\left(h\right)\right]=\llb\lfloor e\rfloor\rrb\left(\left\{ hd\,h\right\} \right)$.
\end{thm}
Again the proof proceeds by induction on $e$, but we will elaborate
slightly this time. The base cases are all straightforward by the
following reasoning. Both $\boldsymbol{0}$ and $\bot$ filter out
all packet (histories), while $\boldsymbol{1}$ and $\top$ let through
everything. Assignments and tests in both worlds perform the same
operations on the (head) packet. $\mathsf{dup}$ does not change the
head packet, and on both sides we only consider the head packets.
The inductive cases then rely on $hd$ commuting with the semantics
of the NetKAT operators $+$, $\cdot$ and $^{*}$, which it does
since NetKAT terms do not examine or modify packets in the packet
history beyond the head.

\section{Equational Theories of MatchKAT and $\mathsf{dup}$-Free NetKAT\label{sec:Equational-Theories-of}}

As promised, we show that the equational theory of MatchKAT is sound
and complete with respect to the packet filtering semantics, through
borrowing soundness and completeness results of NetKAT's equational
theory from \cite{Anderson:2014:NSF:2535838.2535862}.

Consider two NetKAT terms $e$ and $e'$. Suppose
\[
\forall h\in\mathsf{H}.\,hd\left[\llb e\rrb\left(h\right)\right]=hd\left[\llb e'\rrb\left(h\right)\right],
\]
it is not necessarily the case that $\llb e\rrb=\llb e'\rrb$. Although
NetKAT terms cannot access packets beyond the head in the input packet
history, $\llb e\rrb$ may still produce different output packet histories
compared to $\llb e'\rrb$ by using $\mathsf{dup}$. If $e$ and $e'$
are $\mathsf{dup}$-free however, we are then able to deduce $\llb e\rrb=\llb e'\rrb$.
The equational theory of $\mathsf{dup}$-free NetKAT is therefore
determined entirely by the operations on the head packet. This idea
can be developed into a proof for the soundness and completeness for
the equational theory of MatchKAT. First we require two lemmas.
\begin{lem}
\label{lem:to-from-lemma}For any MatchKAT expression $e$, $\llb e\rrb=\llb\lfloor\lceil e\rceil\rfloor\rrb$.
\end{lem}
\begin{lem}
\label{lem:to-preserves-equiv}For all MatchKAT expressions $e$ and
$e'$, $e\equiv e'\iff\lceil e\rceil\equiv\lceil e'\rceil$.
\end{lem}
Proofs of these results can be found in the Appendix, with the insight
in both being that translations to/from NetKAT preserve equations
syntactically and semantically.
\begin{thm}
\label{thm:(Soundness-and-completeness.)}(Soundness and completeness.)
For all MatchKAT expressions $e$ and $e'$, $e\equiv e'\iff\llb e\rrb=\llb e'\rrb$.
\end{thm}
This follows from the implications
\[
\begin{array}{rcll}
e\equiv e' & \iff & \lceil e\rceil\equiv\lceil e'\rceil & \text{(Lemma \ref{lem:to-preserves-equiv})}\\
 & \iff & \llb\lceil e\rceil\rrb=\llb\lceil e'\rceil\rrb & \text{(NetKAT sound \& completeness)}\\
 & \iff & \llb\lfloor\lceil e\rceil\rfloor\rrb=\llb\lfloor\lceil e'\rceil\rfloor\rrb & \text{(Theorem \ref{thm:netkat-to-matchkat})}\\
 & \iff & \llb e\rrb=\llb e'\rrb & \text{(Lemma \ref{lem:to-from-lemma})}.
\end{array}
\]
This third step follows from Theorem $\ref{thm:netkat-to-matchkat}$
since $\lceil e\rceil$ and $\lceil e'\rceil$, being translations
from MatchKAT and therefore $\mathsf{dup}$-free, have interpretations
determined entirely by modifications on the head packet.

\subsection{Complexity of Deciding Equivalence\label{subsec:Complexity-of-Deciding}}

In this section, we discuss the complexity of deciding equivalence
in MatchKAT, and how the result relates to NetKAT.
\begin{thm}
\label{thm:Deciding-equivalence-in}Deciding equivalence in MatchKAT
is $\mathsf{PSPACE}$-complete
\end{thm}
Membership of $\mathsf{PSPACE}$ is argued by translating the MatchKAT
terms to the $\mathsf{dup}$-free fragment of NetKAT as shown previously.
The equational theory of this fragment is in $\mathsf{PSPACE}$ since
that of NetKAT is in $\mathsf{PSPACE}$ \cite{Anderson:2014:NSF:2535838.2535862}.

For hardness, we can encode a word problem for a linear-bounded automaton
as a MatchKAT term $e$, such that the automaton accepts the given
word if and only if $e\not\equiv\bot$. The proof is given in the
Appendix. The word problem for a linear-bounded automaton is known
to be $\mathsf{PSPACE}$-hard \cite{Garey:1990:CIG:574848}.

The hardness result in \cite{Anderson:2014:NSF:2535838.2535862} of
deciding equivalence in NetKAT relies on a simple translation of regular
expressions to NetKAT expressions containing many $\mathsf{dup}$s.
Our result improves this slightly:
\begin{cor}
Deciding equivalence of $\mathsf{dup}$-free NetKAT terms is $\mathsf{PSPACE}$-complete.
\end{cor}
This can be seen through a similar encoding of the linear-bounded
automaton.

\section{Discussion and Conclusion\label{sec:Discussion}}

We will end by discussing the potential applications and decision
procedures of MatchKAT, as the latter will be crucial in any real-world
application in reasoning with match-action tables. Efficient procedures
for NetKAT have already been discovered, such as in \cite{Foster:2015:CDP:2676726.2677011,Smolka:2015:FCN:2784731.2784761},
that work well on many real-life cases. Much of the difficult work
is in reasoning with $\mathsf{dup}$, which MatchKAT does without.
We conjecture that it should be possible to adapt these previous decision
procedures to MatchKAT with much simplification. A coalgebraic treatment
of MatchKAT directly is also conjectured to be possible.

Application-wise, it is envisaged that MatchKAT could be used to reason
about local switch behavior, in contrast to NetKAT on global network
policies, when the switch has already been configured by match-action
rules. This could be useful for various reasons:
\begin{itemize}
\item MatchKAT has a sound and complete equational theory. Equivalence of
terms can be decided and is guaranteed to be sound. This helps in
the verification of correctness as well as potential configuration
optimizations in reducing the number of rules. We have previously
talked about the notion of length for MatchKAT terms, and so equivalence
of terms of different lengths is potentially proof of equivalence
between optimized and unoptimized configurations.
\item MatchKAT is equivalent to $\mathsf{dup}$-free NetKAT, and there is
a well-defined translation between the two. This could help in decompiling
match-action tables to NetKAT in order to make sense of the global
policies they are implementing.
\item MatchKAT's match expressions is closer to how bits in packet headers
are matched on switches at low-level. MatchKAT could potentially help
with efficient implementations of hardware that performs matching.
\end{itemize}
These all distinguish our work from previous attempts such as \cite{KazemianPhD,180587}
that also reasoned with binary data in packet headers theoretically.
We also note with interest that other authors have also created new
algebraic systems with a strong relationship to NetKAT, such as \cite{10.1007/978-3-030-02149-8_17}.
In the future, we intend to further develop concrete applications
of MatchKAT in the setting of match-action tables, and demonstrate
the usefulness of its algebraic theory.\bibliographystyle{plainurl}
\phantomsection\addcontentsline{toc}{section}{\refname}\bibliography{matchkat}

\begin{thebibliography}{10}

\bibitem{Anderson:2014:NSF:2535838.2535862}
Carolyn~Jane Anderson, Nate Foster, Arjun Guha, Jean-Baptiste Jeannin, Dexter
  Kozen, Cole Schlesinger, and David Walker.
\newblock {NetKAT}: Semantic foundations for networks.
\newblock In {\em Proceedings of the 41st ACM SIGPLAN-SIGACT Symposium on
  Principles of Programming Languages}, POPL '14, pages 113--126, New York, NY,
  USA, 2014. ACM.
\newblock URL: \url{http://doi.acm.org/10.1145/2535838.2535862}, \href
  {https://doi.org/10.1145/2535838.2535862}
  {\path{doi:10.1145/2535838.2535862}}.

\bibitem{Bosshart:2014:PPP:2656877.2656890}
Pat Bosshart, Dan Daly, Glen Gibb, Martin Izzard, Nick McKeown, Jennifer
  Rexford, Cole Schlesinger, Dan Talayco, Amin Vahdat, George Varghese, and
  David Walker.
\newblock P4: Programming protocol-independent packet processors.
\newblock {\em SIGCOMM Comput. Commun. Rev.}, 44(3):87--95, July 2014.
\newblock URL: \url{http://doi.acm.org/10.1145/2656877.2656890}, \href
  {https://doi.org/10.1145/2656877.2656890}
  {\path{doi:10.1145/2656877.2656890}}.

\bibitem{Choi:2017:PPV:3050220.3060609}
Sean Choi, Xiang Long, Muhammad Shahbaz, Skip Booth, Andy Keep, John Marshall,
  and Changhoon Kim.
\newblock Pvpp: A programmable vector packet processor.
\newblock In {\em Proceedings of the Symposium on SDN Research}, SOSR '17,
  pages 197--198, New York, NY, USA, 2017. ACM.
\newblock URL: \url{http://doi.acm.org/10.1145/3050220.3060609}, \href
  {https://doi.org/10.1145/3050220.3060609}
  {\path{doi:10.1145/3050220.3060609}}.

\bibitem{conway1971regular}
J.H. Conway.
\newblock {\em Regular algebra and finite machines}.
\newblock Chapman and Hall mathematics series. Chapman and Hall, 1971.
\newblock URL: \url{https://books.google.com/books?id=xBXvAAAAMAAJ}.

\bibitem{Foster:2015:CDP:2676726.2677011}
Nate Foster, Dexter Kozen, Matthew Milano, Alexandra Silva, and Laure Thompson.
\newblock A coalgebraic decision procedure for {NetKAT}.
\newblock In {\em Proceedings of the 42Nd Annual ACM SIGPLAN-SIGACT Symposium
  on Principles of Programming Languages}, POPL '15, pages 343--355, New York,
  NY, USA, 2015. ACM.
\newblock URL: \url{http://doi.acm.org/10.1145/2676726.2677011}, \href
  {https://doi.org/10.1145/2676726.2677011}
  {\path{doi:10.1145/2676726.2677011}}.

\bibitem{Garey:1990:CIG:574848}
Michael~R. Garey and David~S. Johnson.
\newblock {\em Computers and Intractability; A Guide to the Theory of
  NP-Completeness}.
\newblock W. H. Freeman \& Co., New York, NY, USA, 1990.

\bibitem{10.1007/978-3-030-02149-8_17}
Malvin Gattinger and Jana Wagemaker.
\newblock Towards an analysis of dynamic gossip in {NetKAT}.
\newblock In Jules Desharnais, Walter Guttmann, and Stef Joosten, editors, {\em
  Relational and Algebraic Methods in Computer Science}, pages 280--297, Cham,
  2018. Springer International Publishing.

\bibitem{KazemianPhD}
Peyman Kazemian.
\newblock {\em Header Space Analysis}.
\newblock PhD thesis, Stanford University, 2013.

\bibitem{180587}
Peyman Kazemian, George Varghese, and Nick McKeown.
\newblock Header space analysis: Static checking for networks.
\newblock In {\em Presented as part of the 9th {USENIX} Symposium on Networked
  Systems Design and Implementation ({NSDI} 12)}, pages 113--126, San Jose, CA,
  2012. {USENIX}.
\newblock URL:
  \url{https://www.usenix.org/conference/nsdi12/technical-sessions/presentation/kazemian}.

\bibitem{Kozen:1997:KAT:256167.256195}
Dexter Kozen.
\newblock Kleene algebra with tests.
\newblock {\em ACM Trans. Program. Lang. Syst.}, 19(3):427--443, May 1997.
\newblock URL: \url{http://doi.acm.org/10.1145/256167.256195}, \href
  {https://doi.org/10.1145/256167.256195} {\path{doi:10.1145/256167.256195}}.

\bibitem{Kozen:1997:AC:549365}
Dexter~C. Kozen.
\newblock {\em Automata and Computability}.
\newblock Springer-Verlag, Berlin, Heidelberg, 1st edition, 1997.

\bibitem{Lakshminarayanan:2005:AAP:1080091.1080115}
Karthik Lakshminarayanan, Anand Rangarajan, and Srinivasan Venkatachary.
\newblock Algorithms for advanced packet classification with ternary {CAM}s.
\newblock In {\em Proceedings of the 2005 Conference on Applications,
  Technologies, Architectures, and Protocols for Computer Communications},
  SIGCOMM '05, pages 193--204, New York, NY, USA, 2005. ACM.
\newblock URL: \url{http://doi.acm.org/10.1145/1080091.1080115}, \href
  {https://doi.org/10.1145/1080091.1080115}
  {\path{doi:10.1145/1080091.1080115}}.

\bibitem{McKeown:2008:OEI:1355734.1355746}
Nick McKeown, Tom Anderson, Hari Balakrishnan, Guru Parulkar, Larry Peterson,
  Jennifer Rexford, Scott Shenker, and Jonathan Turner.
\newblock Openflow: Enabling innovation in campus networks.
\newblock {\em SIGCOMM Comput. Commun. Rev.}, 38(2):69--74, March 2008.
\newblock URL: \url{http://doi.acm.org/10.1145/1355734.1355746}, \href
  {https://doi.org/10.1145/1355734.1355746}
  {\path{doi:10.1145/1355734.1355746}}.

\bibitem{Shahbaz:2016:PPP:2934872.2934886}
Muhammad Shahbaz, Sean Choi, Ben Pfaff, Changhoon Kim, Nick Feamster, Nick
  McKeown, and Jennifer Rexford.
\newblock Pisces: A programmable, protocol-independent software switch.
\newblock In {\em Proceedings of the 2016 ACM SIGCOMM Conference}, SIGCOMM '16,
  pages 525--538, New York, NY, USA, 2016. ACM.
\newblock URL: \url{http://doi.acm.org/10.1145/2934872.2934886}, \href
  {https://doi.org/10.1145/2934872.2934886}
  {\path{doi:10.1145/2934872.2934886}}.

\bibitem{Smolka:2015:FCN:2784731.2784761}
Steffen Smolka, Spiridon Eliopoulos, Nate Foster, and Arjun Guha.
\newblock A fast compiler for {NetKAT}.
\newblock In {\em Proceedings of the 20th ACM SIGPLAN International Conference
  on Functional Programming}, ICFP 2015, pages 328--341, New York, NY, USA,
  2015. ACM.
\newblock URL: \url{http://doi.acm.org/10.1145/2784731.2784761}, \href
  {https://doi.org/10.1145/2784731.2784761}
  {\path{doi:10.1145/2784731.2784761}}.

\end{thebibliography}

\appendix

\section*{Appendix}

\subsection*{Proof of Lemma \ref{lem:to-from-lemma}}

For all $h$, we have 
\[
hd\left[\llb\lceil e\rceil\rrb\left(h\right)\right]=\llb\lfloor\lceil e\rceil\rfloor\rrb\left(\left\{ hd\,h\right\} \right)
\]
by Theorem \ref{thm:netkat-to-matchkat}. Hence for all $H\subseteq\mathsf{H}$,
\[
\bigcup_{h\in H}hd\left[\llb\lceil e\rceil\rrb\left(h\right)\right]=\llb\lfloor\lceil e\rceil\rfloor\rrb\left(hd\,H\right)
\]
On the other hand, Theorem \ref{thm:matchkat-to-netkat} gives us
\[
\llb e\rrb\left(hd\,H\right)=\bigcup_{\pi\in hd\,H}hd\left[\llb\lceil e\rceil\rrb\left(\pi::\left\langle \right\rangle \right)\right]
\]
for all $H\subseteq\mathsf{H}$. Since $h\approx\left(hd\,h\right)::\left\langle \right\rangle $
for all $h\in H$, it is safe to rewrite the latter equation to
\[
\llb e\rrb\left(hd\,H\right)=\bigcup_{h\in H}hd\left[\llb\lceil e\rceil\rrb\left(h\right)\right].
\]
Combining this with the second equation gives the required result.

\subsection*{Proof of Lemma \ref{lem:to-preserves-equiv}}

\[
e\equiv e'\iff\lceil e\rceil\equiv\lceil e'\rceil
\]
On the left we have MatchKAT terms operating on $n$ bits. On the
right are NetKAT terms operating on $n$ fields each containing $1$
bit. Through exhaustion we can prove that every axiom in the MatchKAT
world gives rise to a corresponding axiom (or derivable theorem) in
the NetKAT world, or vice versa, and hence a proof of equality in
one produces automatically a proof of equality in the other. Instead
of going through the full proof for every axiom, we give some reasons
for why it works.
\begin{itemize}
\item KAT axioms are clearly present in both worlds, and $\lceil\cdot\rceil$
is a homomorphism.
\item The packet algebra axioms are present in both as mentioned in Section
\ref{subsec:matchkat-Definitions}.
\item The axioms for manipulating match expressions are present in MatchKAT,
but they are not in NetKAT. However NetKAT has extra axioms of the
forms $i=k\cdot i=k\equiv i=k$, $k\neq k'\implies i=k\cdot i=k'\equiv\boldsymbol{0}$,
and $\sum_{k}\left(i=k\right)\equiv\boldsymbol{1}$. These, along
with the axioms of the Boolean algebra, are sufficient to derive equivalents
of match expression axioms as theorems.
\end{itemize}

\subsection*{Proof of Theorem \ref{thm:Deciding-equivalence-in}}

A linear-bounded automaton $M=\left(Q,\Sigma,\vdash,\dashv,\delta,s,t,r\right)$
is composed of:
\begin{itemize}
\item Finite set of states $Q$.
\item Tape alphabet $\Sigma$.
\item Left $\vdash$ and right $\dashv$ tape-end markers.
\item Transition relation $\delta\subseteq\left[Q\times\left(\Sigma\cup\left\{ \vdash,\dashv\right\} \right)\right]\times\left[Q\times\left(\Sigma\cup\left\{ \vdash,\dashv\right\} \right)\times\left\{ L,R\right\} \right]$.
\item Start $s$, accept $t$, and reject $r$ states.
\end{itemize}
$M$ can be seen as a non-deterministic Turing machine where the tape
is finite and marked on both ends by $\vdash$ and $\dashv$. At the
start, an input word is present on the tape, while the tape is bound
to a linear size $n$ with respect to the length of the input. $\delta$
is restricted such that the end markers are unmodified and the tape
head does not move off the ends of the tape. The automaton never transitions
out of the accept or reject states once it enters them. We will further
restrict $\Sigma$ to two symbols $\left\{ 0,1\right\} $. This is
without loss of generality with a linear increase in the amount of
tape required.

A packet in the MatchKAT encoding of $M$ contains the following bits:
\begin{itemize}
\item $n$ bits that we will refer to by convience as $\mathsf{tape}_{1},\dots,\mathsf{tape}_{n}$,
representing the tape.
\item Bits $\mathsf{state}_{1},\dots,\mathsf{state}_{\log\left|Q\right|}$
to record the binary encoding of the current state.
\item Bits $\mathsf{head}_{1},\dots,\mathsf{head}_{\log n}$ to record the
binary encoding of the head position.
\end{itemize}
Instead of referring to each state and head bit individually, we will
assign and test for them collectively for a particular $Q$ state
or tape head position. We then construct expressions in the MatchKAT
as follows:
\begin{itemize}
\item The \emph{setup expression }$\alpha$, which is an assignment of $\mathsf{state}$
with $s$, $\mathsf{head}$ with $1$, and the $\mathsf{tape}$ fields
as appropriate for the initial tape contents for a given input word..
\item The \emph{transition expression }$\beta$, consisting of sums guarded
by $\mathsf{state}\times\mathsf{head}\times\mathsf{tape}_{\mathsf{head}}$
conditions. For a packet in a given configuration, it rewrites it
as per one action of the transition relation.
\item The \emph{decision expression} $\gamma$, which is just a test for
$\mathsf{state}=t$.
\end{itemize}
Consider the expression $\alpha\left(\beta^{*}\right)\gamma$, which
is not equivalent to $\bot$ if and only $M$ accepts the given word.
For any non-empty set of input packets, $\alpha\left(\beta^{*}\right)$
constructs the set of all reachable configurations of $M$, while
$\gamma$ filters this set to include only the packets that contain
the accept state. On the other hand, $\bot$ drops all packets. The
size of the expressions are polynomial in the size of the automaton
specification.
\end{document}